\documentclass[a4paper,11pt]{article}
\usepackage{pos}

\begin{document}

\title{Boson-jet and jet-jet azimuthal correlations at high transverse momenta}
\ShortTitle{Boson-jet and jet-jet azimuthal correlations}

\author*[a]{A.M.~van Kampen}
\author[b]{A.~Bermudez Martinez}
\author[b]{L.I.~Estevez Banos}
\author[a,c,d]{F.~Hautmann}
\author[b]{H.~Jung}
\author[b]{M.~Mendizabal}
\author[e]{K.~Moral Figueroa}
\author[f]{S.~Prestel}
\author[b]{S.~Taheri Monfared}
\author[b,g]{Q.~Wang}
\author[b]{K.~Wichmann} 
\author[b,g]{H.~Yang}

\affiliation[a]{Elementary Particle Physics, University of Antwerp, Antwerp, Belgium}
\affiliation[b]{Deutsches Elektronen-Synchrotron DESY, Hamburg, Germany}
\affiliation[c]{CERN, Geneva, Switzerland}
\affiliation[d]{University of Oxford, Oxford, UK}
\affiliation[e]{University of Edinburgh, Edinburgh, UK}
\affiliation[f]{Department of Astronomy and Theoretical Physics, Lund University, Lund, Sweden}
\affiliation[g]{School of Physics, Peking University, Beijing, China}

\emailAdd{AronMees.vanKampen@uantwerpen.be}

\abstract{We discuss our recent results on azimuthal distributions in 
vector boson + jets and multi-jet production at the LHC, obtained from the matching of 
next-to-leading order (NLO) perturbative matrix elements with 
 transverse momentum dependent (TMD) parton branching. We present a comparative analysis of 
 boson-jet and jet-jet correlations in the back-to-back region, and a study of the 
 theoretical systematic uncertainties associated with the matching scale in the cases  of 
 TMD and collinear parton showers.} 

\FullConference{%
  41st International Conference on High Energy physics - ICHEP2022\\
  6-13 July, 2022\\
  Bologna, Italy
}

\maketitle

Experiments at the Large Hadron Collider (LHC) 
carry out accurate measurements of azimuthal correlations 
in vector-boson plus jets~\cite{CMS:2016php,CMS:2018mdf}  
and  
multi-jet~\cite{daCosta:2011ni,Khachatryan:2011zj,Khachatryan:2016hkr,CMS:2017cfb,CMS:2019joc}  
final states.  
When a boson and a jet, or two jets, recoil nearly back-to-back, reliable QCD predictions  call for 
soft-gluon resummation of Sudakov processes --- see recent 
studies of Refs.~\cite{Bouaziz:2022vp,Chien:2022wiq,Buonocore:2021akg,Chien:2020hzh,Chien:2019gyf,Buffing:2018ggv,Sun:2018icb} in the 
boson-jet case and 
Refs.~\cite{Sun:2014gfa,Sun:2015doa,Hatta:2021jcd,Hatta:2020bgy} in the di-jet case.  
This region 
 probes the transverse momentum dependent 
(TMD)~\cite{Abdulov:2021ivr,Hautmann:2014kza,Angeles-Martinez:2015sea}   parton distribution
 of the initial state. 

With the increase in luminosity at the LHC, it becomes possible to  explore this region  experimentally 
 over a wide kinematic range in the hard scale of the process, set by the highest 
transverse momentum $p_T^{\rm leading}$ produced into the final state, from 
$p_T^{\rm leading} \approx {\cal O}$(100 GeV)  to $p_T^{\rm leading} \approx {\cal O}$(1000 GeV). 
 In particular, at the highest 
scales the nearly back-to-back region accessible with the experimental angular resolution 
of about 1 degree is characterized by transverse momentum imbalances of a few ten GeV, 
which can be investigated by analyzing jets with measurable transverse momenta. 

The combined study of the $p_T^{\rm leading}$ and $p_T$-imbalance dependence of TMD dynamics 
is especially important, because the production of colored states near the back-to-back region 
may be influenced by factorization-breaking 
effects~\cite{Rogers:2010dm,Rogers:2013zha,Collins:2007nk,Vogelsang:2007jk}, 
due to interferences of gluon exchange in the soft region~\cite{Collins:1999dz} with 
collinear radiation.   

The present article is based on our work~\cite{Yang:2022qgk}, in which 
we perform studies of  azimuthal correlations  as a function 
of  $p_T^{\rm leading}$,  enabling one to explore 
$p_T$ imbalances from a jet scale of several ten GeV down to the few GeV scale, 
and   propose systematic    measurements of ratios of boson-jet to jet-jet distributions, 
 for varying transverse momenta, 
to investigate potential effects of soft-gluon interferences.

The studies~\cite{Yang:2022qgk} employ the 
parton branching (PB) approach~\cite{Hautmann:2017xtx,Hautmann:2017fcj}        
to TMD evolution, and its 
matching~\cite{BermudezMartinez:2019anj,BermudezMartinez:2020tys,Abdulhamid:2021xtt} to 
next-to-leading-order (NLO) matrix-element calculations.    
The PB TMD evolution and corresponding parton shower are 
implemented in the Monte Carlo event generator {\scshape Cascade3}~\cite{Baranov:2021uol}, 
and the NLO matching is performed using  {\scshape MadGraph5\_aMC@NLO}~\cite{Alwall:2014hca} (we label 
this calculational framework as MCatNLO+CAS3 in the following). 
TMD parton distributions at the starting scale of evolution are obtained from 
fits~\cite{BermudezMartinez:2018fsv} 
to  precision deep-inelastic scattering data~\cite{Abramowicz:2015mha} using the 
xFitter analysis framework~\cite{xFitterDevelopersTeam:2022koz,Alekhin:2014irh}. 

The comparison of  NLO  PB-TMD predictions and collinear-shower predictions with  the 
measurements of  di-jet azimuthal correlations~\cite{CMS:2017cfb,CMS:2019joc} is discussed 
in Refs.~\cite{Abdulhamid:2021xtt,Martinez:2022dux}. The description of jet correlation data 
by NLO  PB-TMD is good, and the 
 comparison   
   underlines the importance of transverse momentum recoils in the parton 
showers~\cite{Dooling:2012uw,Hautmann:2013fla,Jung:2010si,Hautmann:2008vd}   
and of angular ordering~\cite{Marchesini:1987cf,Catani:1990rr,Hautmann:2019biw} in  
achieving this.  For the boson-jet case, the  measurements performed so 
far do not yet have   TeV-scale transverse momenta  
and sufficiently fine binning to investigate detailed QCD features.

\begin{figure}[h!tb]
\begin{center} 
\includegraphics[width=0.45\textwidth]{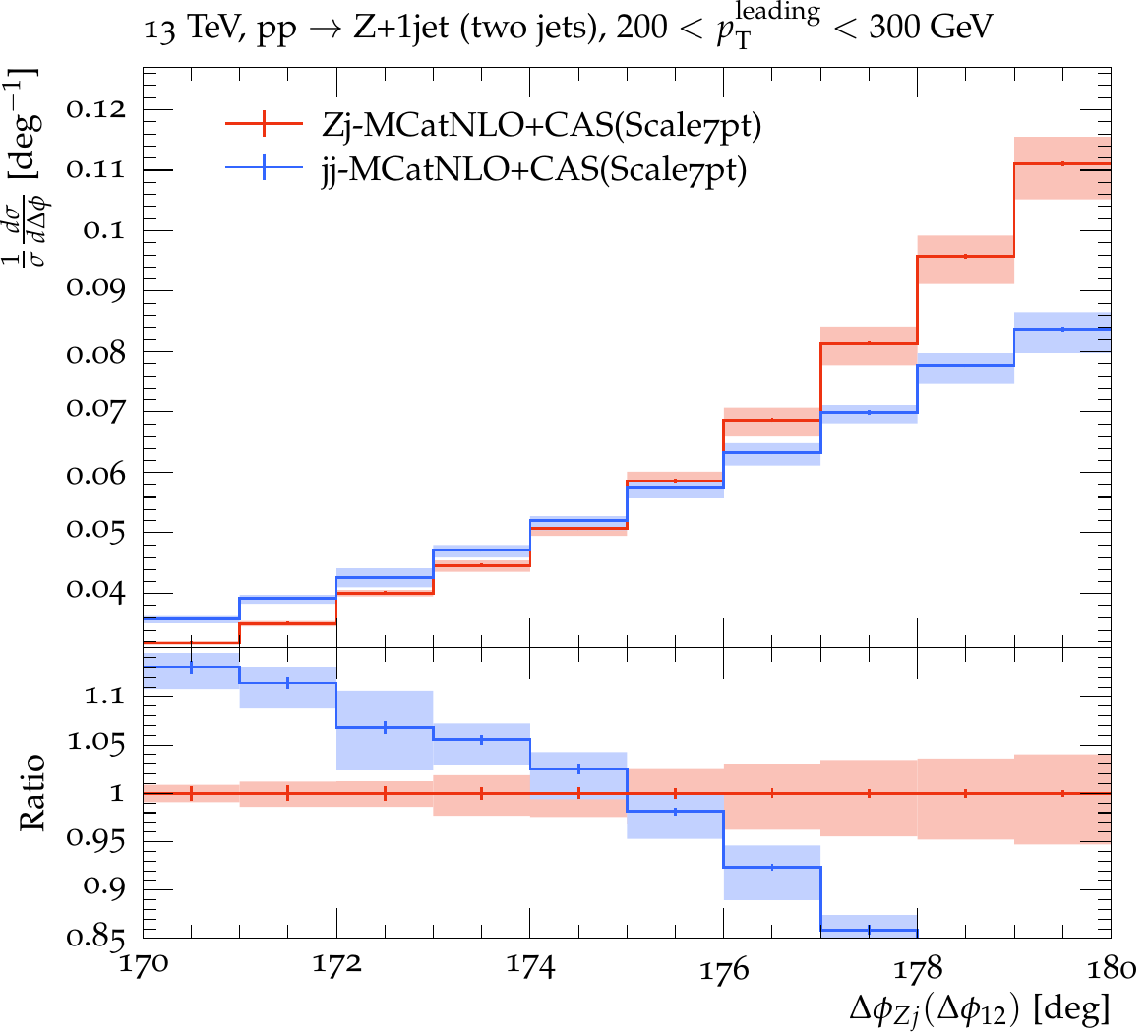} 
\includegraphics[width=0.45\textwidth]{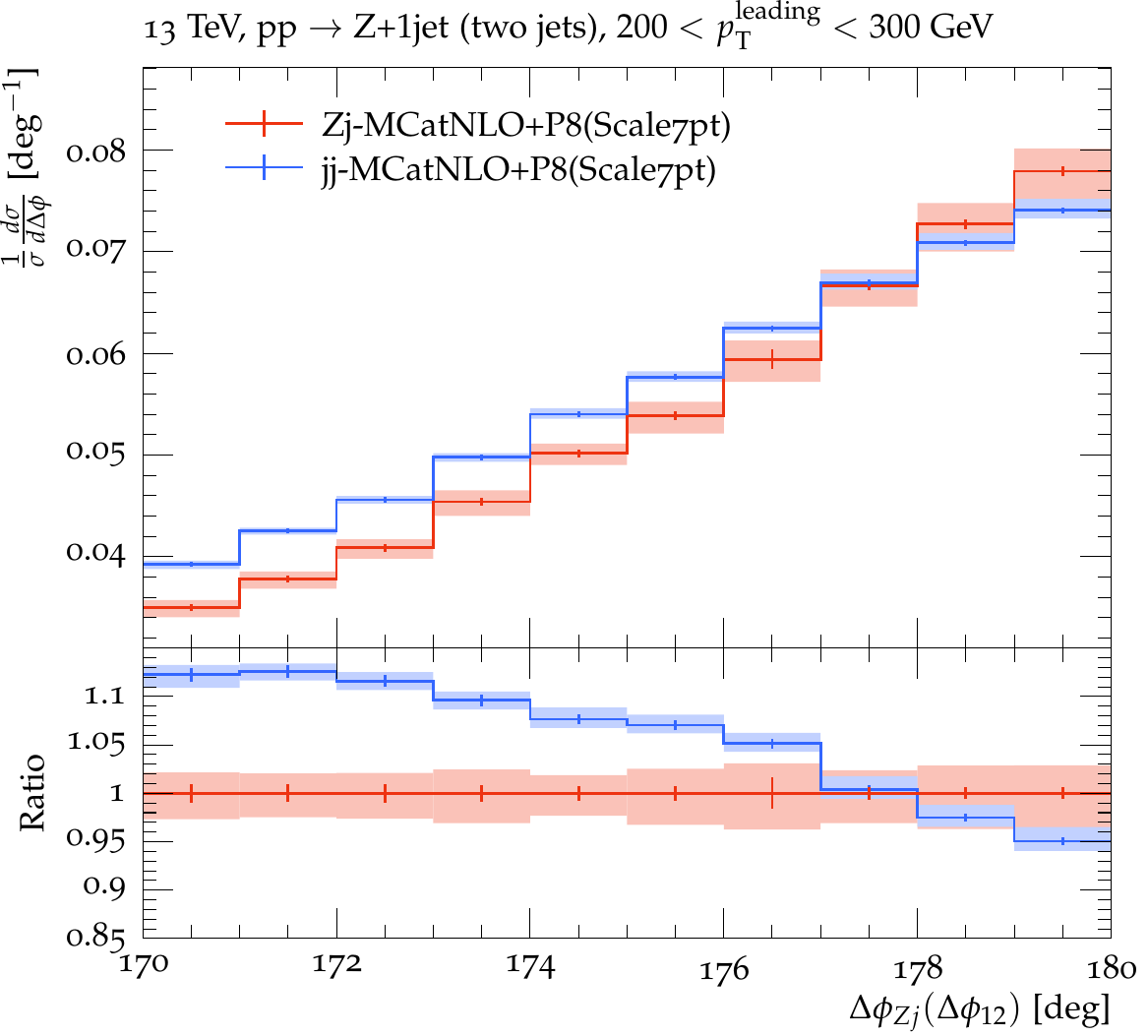} 
\includegraphics[width=0.45\textwidth]{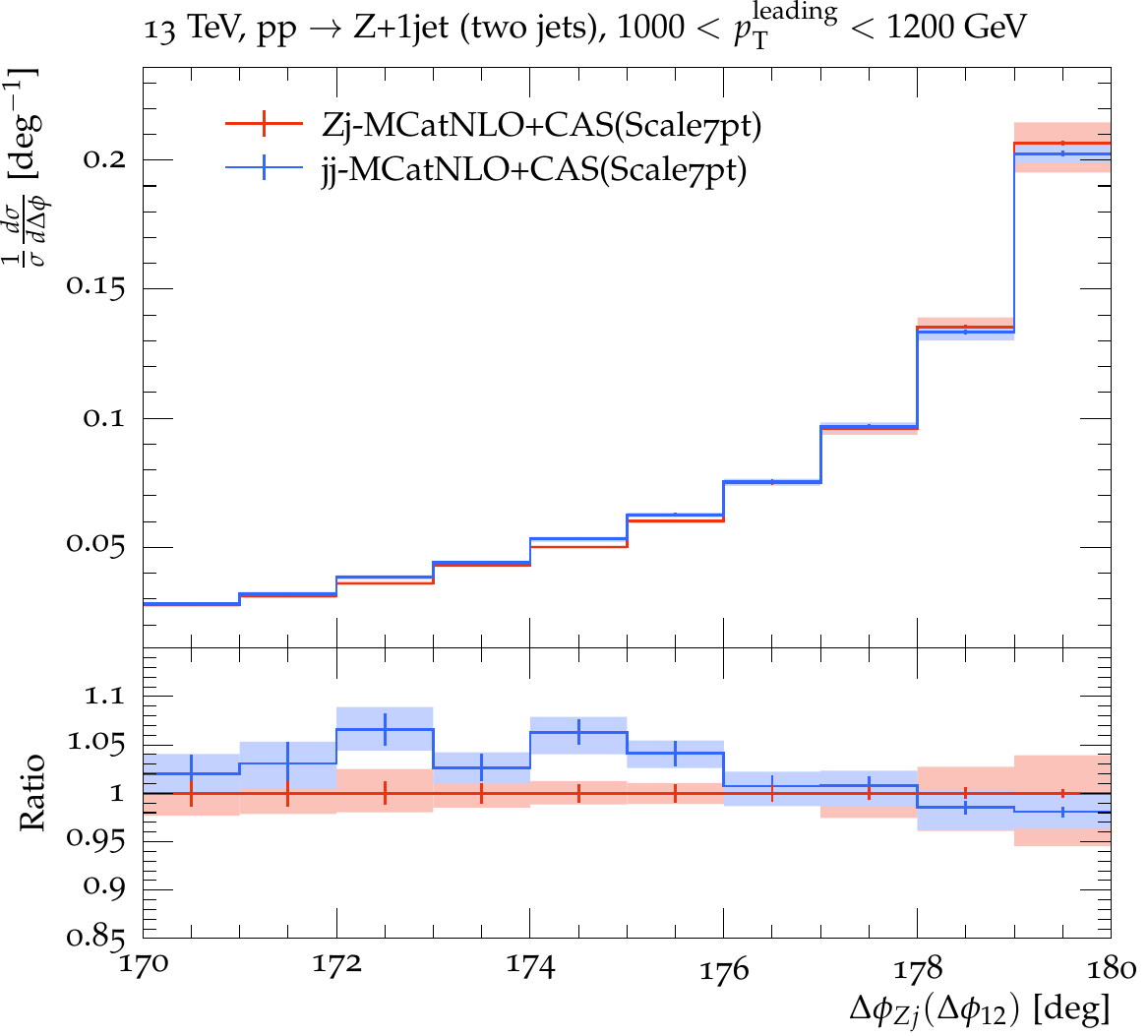} 
\includegraphics[width=0.45\textwidth]{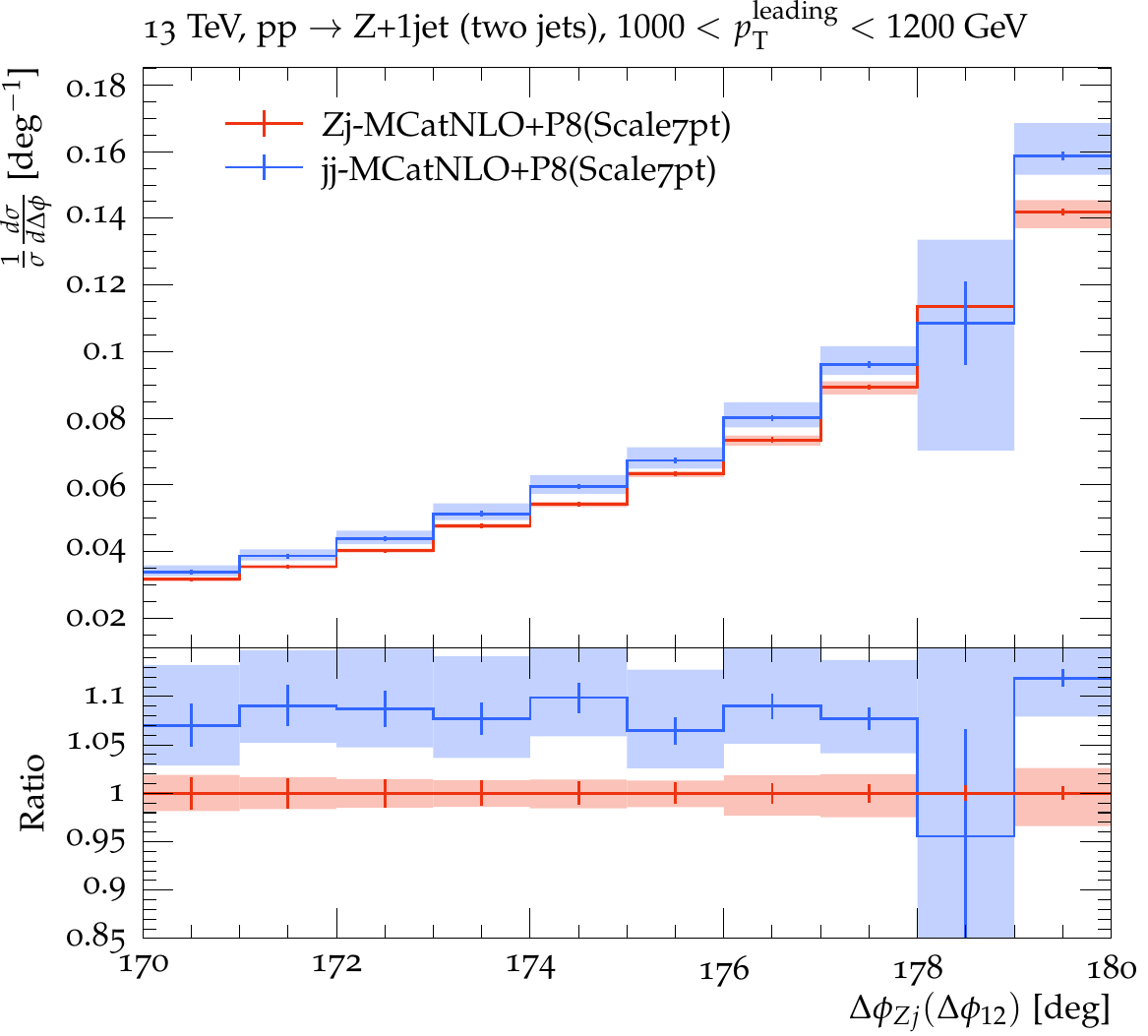} 
\caption{\small Predictions for the azimuthal correlation 
  in the back-to-back region for Z+jets  and multijet production obtained with MCatNLO+CAS3 (left column) and  
 MCatNLO+{\scshape Pythia8} (right column)~\protect\cite{Yang:2022qgk}. 
 Shown are different regions in  $p_T > 200 $ GeV (upper row) and 
 $p_T > 1000 $ GeV  (lower row). The bands give an estimate of  theoretical 
  uncertainties obtained from scale variations as described in~\protect\cite{Yang:2022qgk}.}
\label{b2b-fig1}
\end{center}
\end{figure} 

Fig.~\ref{b2b-fig1}~\cite{Yang:2022qgk} shows NLO-matched 
predictions from TMD shower (left) and collinear shower~\cite{Sjostrand:2014zea} (right),  
 illustrating spectra  in the $\Delta \phi$  azimuthal separation 
 in Z + jet and di-jet systems. 
 We  concentrate on the large-$\Delta \phi$ region, as the low-$\Delta \phi$ decorrelation 
 region requires going beyond the framework of the present calculation to include 
 the contributions of higher jet multiplicities, e.g.~via multi-jet merging 
 techniques~\cite{Martinez:2021chk,Martinez:2022wrf}. For 
 low $p_T^{\rm leading}$  the boson-jet final state is more 
strongly correlated azimuthally than the jet-jet final state (top panels). When the  
transverse momenta increase, 
 the boson-jet and jet-jet states become more 
 similarly correlated (bottom panels). One can 
 connect  this behavior  to features of the partonic 
 initial state and final state radiation  in the two cases~\cite{Yang:2022qgk}. 
Since  potential factorization-breaking effects arise from 
 color interferences of initial-state and final-state radiation, 
 different  breaking patterns can be expected for  
 strong and weak azimuthal correlations, 
 influencing differently the boson-jet and jet-jet cases.  Thus we  propose 
 to  compare measurements of 
 di-jet  and Z + jet distributions systematically, scanning the 
 phase space  from low 
 to high $p_T^{\rm leading}$. 

\begin{figure}[h!tb]
\begin{center} 
\includegraphics[width=0.45\textwidth]{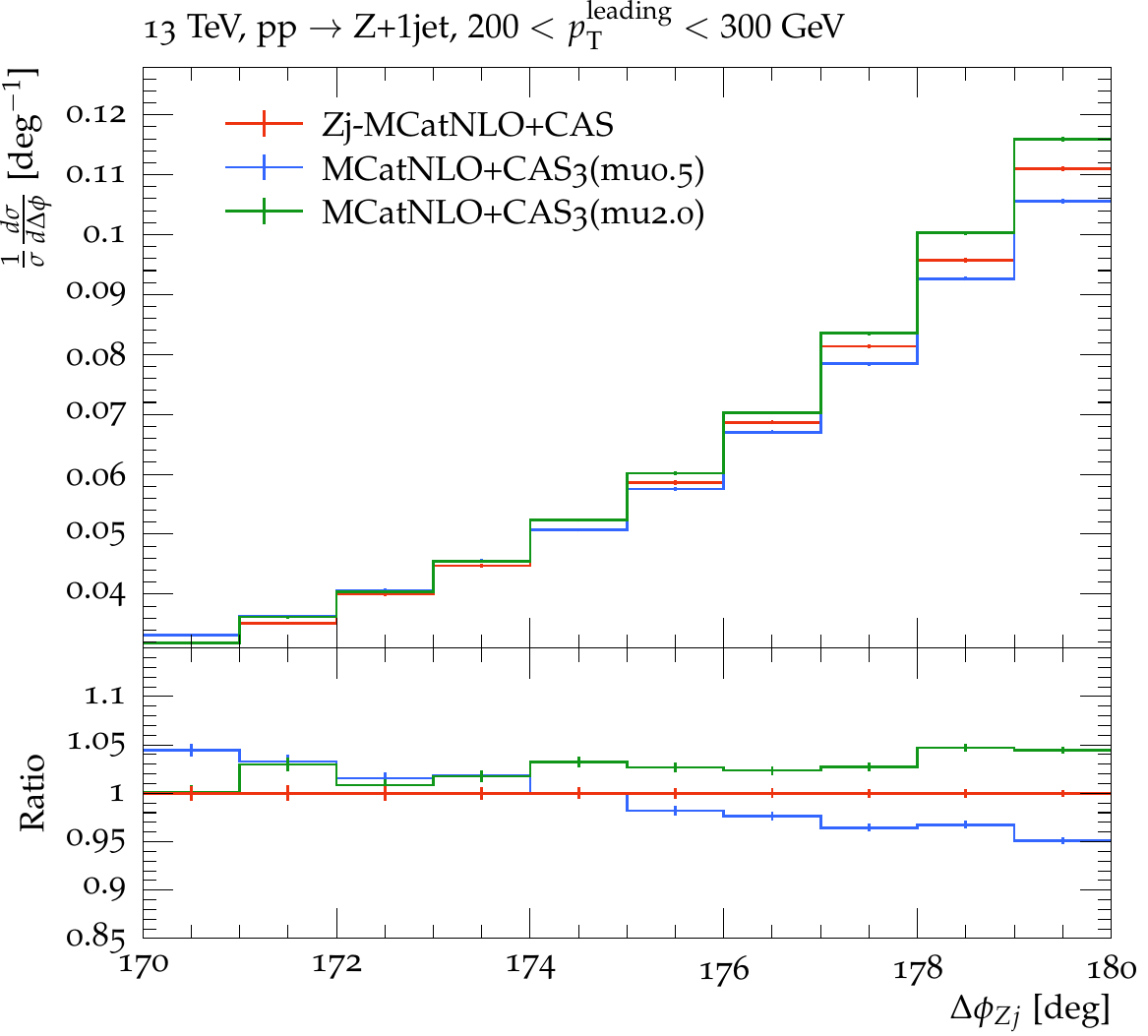} 
\includegraphics[width=0.45\textwidth]{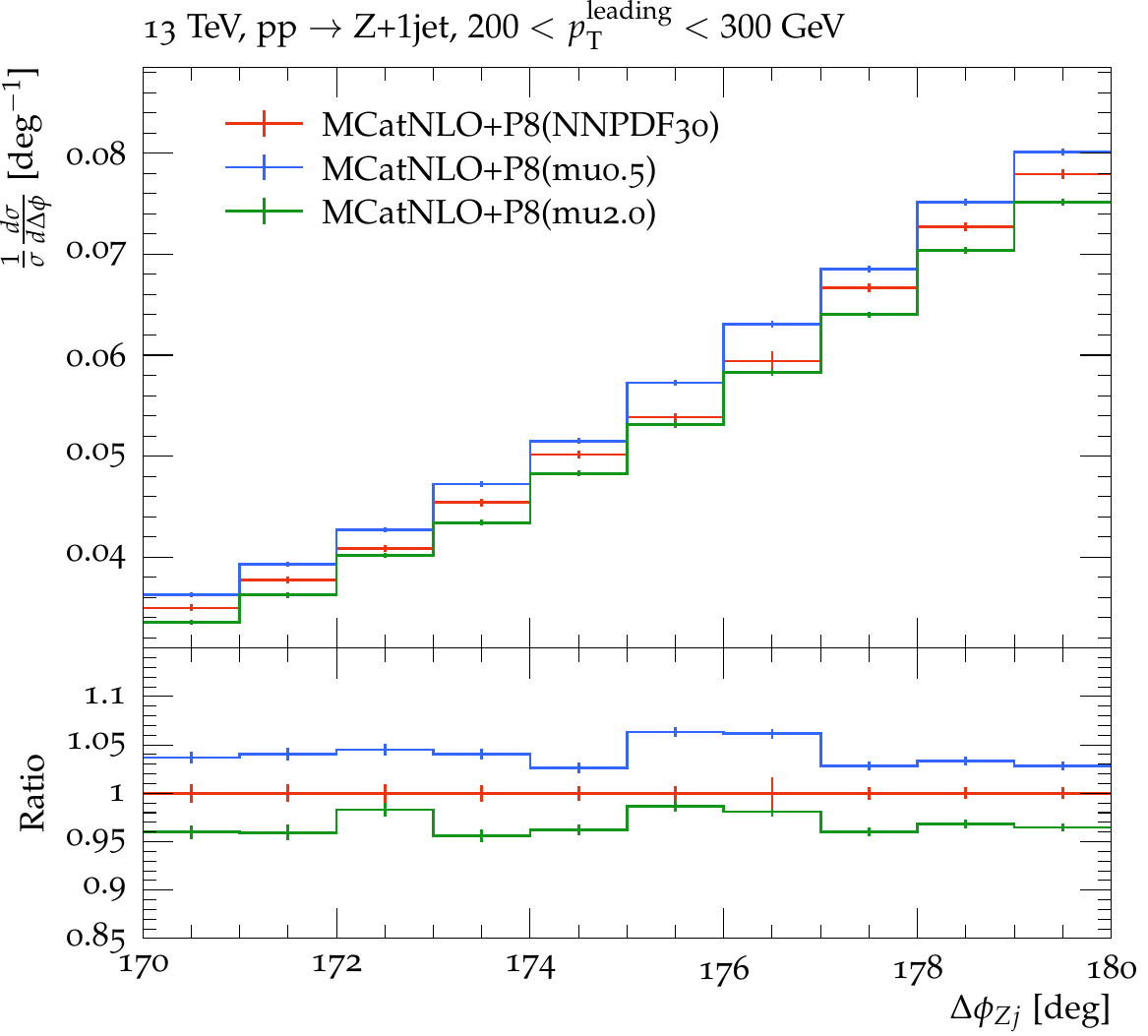} 
\includegraphics[width=0.45\textwidth]{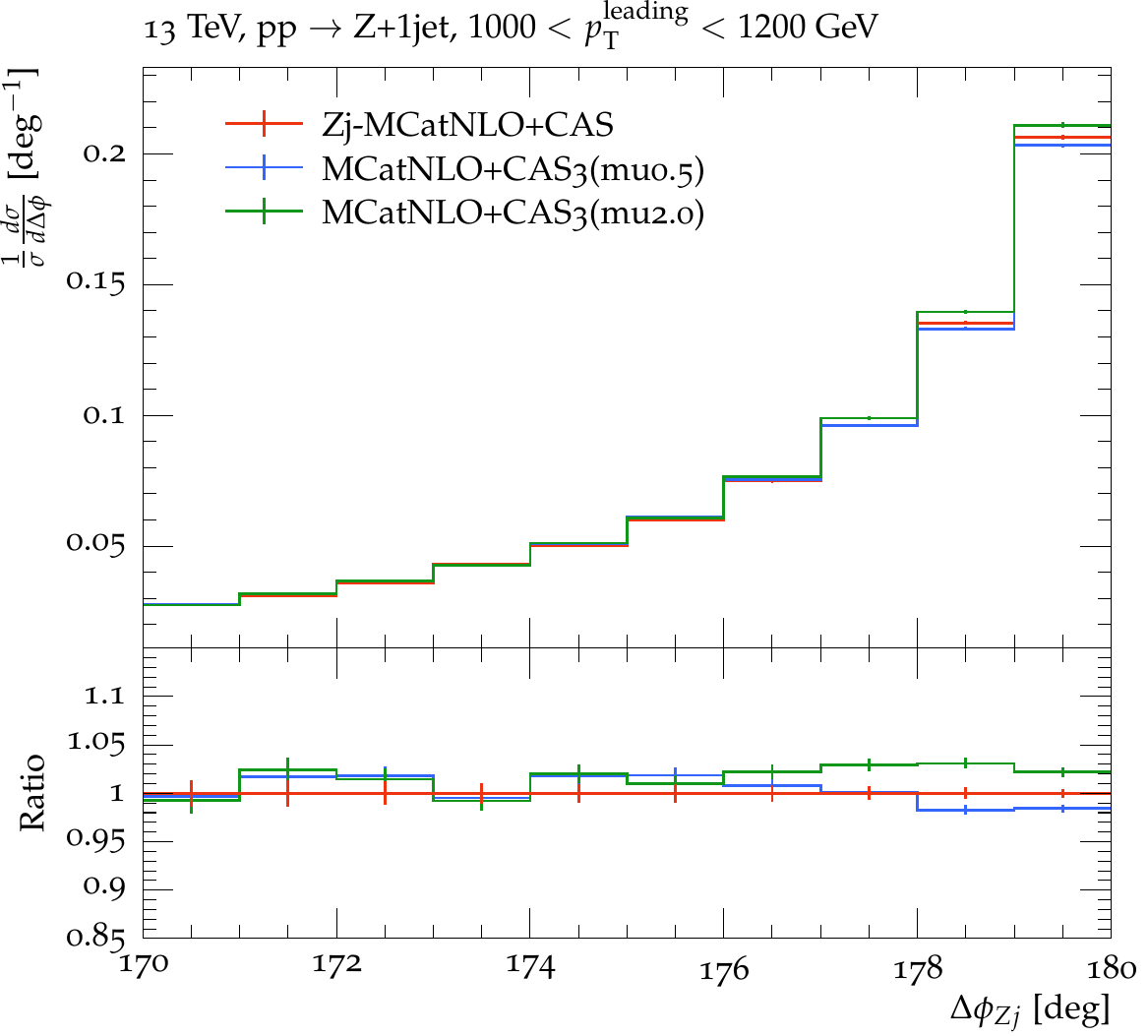} 
\includegraphics[width=0.45\textwidth]{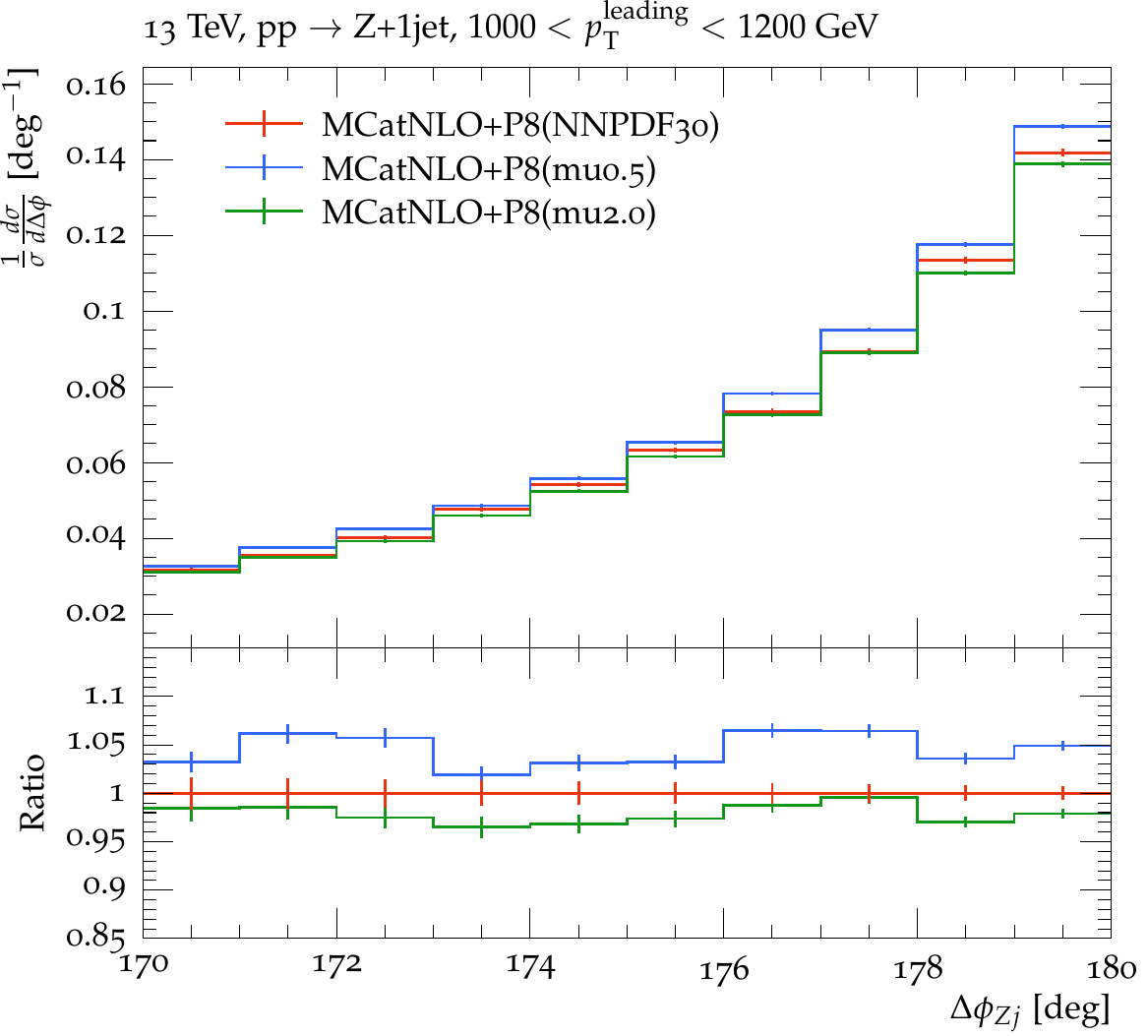} 
  \caption{\small The dependence on the variation of the matching scale $\mu_{m}$ in
  predictions for the azimuthal correlation  in the back-to-back region~\protect\cite{Yang:2022qgk}.
  Shown are predictions  obtained with MCatNLO+CAS3 (left column) and  
    MCatNLO+{\scshape Pythia8} (right column)
  for $p_T > 200 $  GeV (upper row) and $p_T > 1000 $    GeV (lower row).  
  The predictions with different matching scales $\mu_{m}$ varied by a factor of two up and down are shown.}
\label{b2b-fig2}
\end{center}
\end{figure} 

An important source of theoretical systematic uncertainties is given by the matching scale 
$\mu_m$, limiting the hardness of parton shower radiation. In Fig.~\ref{b2b-fig2} we study this 
theoretical systematics for TMD shower (left) and collinear shower (right) calculations. 
Variations of the matching scale  
lead to more stable predictions in the TMD case, with  the 
relative reduction of the matching scale 
theoretical uncertainty becoming more pronounced for increasing 
transverse momenta~\cite{Yang:2022qgk}. 

In conclusion, azimuthal distributions provide 
useful observables to 
gain insight into TMD dynamics and 
factorization ---  see e.g.~\cite{Chien:2022wiq,Abdulhamid:2021xtt}.  
By varying the leading transverse momentum,  
 the relative influence on 
the $p_T$ imbalance can be studied from 
perturbative components and 
non-perturbative effects, e.g.~\cite{Bury:2022czx,Hautmann:2020cyp}. 
Sensitivity to TMD splitting probabilities~\cite{Hautmann:2022xuc}  
could also be observed through azimuthal correlations. 

\vskip 0.3 cm 

\noindent {\bf Acknowledgments}. We thank the conference organizers and 
convenors for putting up a great conference and for the invitation.

\end{document}